\documentclass[12pt,twoside]{article}   
\usepackage[super,sort,comma]{natbib}

\usepackage{fancyhdr}		
\newcommand{\captionv}[3]{\begin{center}\parbox{#1cm}{\caption[#2]{{\sf #3}}}
\end{center}}

\usepackage[section]{placeins}   %

\usepackage{graphicx}
\usepackage{tabularray}
\usepackage{multirow}

\makeatletter \renewcommand\@biblabel[1]{$^{#1}$} \makeatother
\newcommand{\cen}[1]{\begin{center} #1 \end{center}}

\usepackage{hyperref}
\hypersetup{ colorlinks,
    citecolor=blue,
    filecolor=blue,
    linkcolor=blue,
    urlcolor=blue
}

\usepackage{xcolor}

\definecolor{gray}{rgb}{0.6,0.6,0.6}
\definecolor{red}{rgb}{0.85,0,0}
\definecolor{green}{rgb}{0,0.85,0}
\definecolor{blue}{rgb}{0,0,0.85}
\definecolor{beige}{rgb}{0.92,0.87,0.78}
\usepackage[all]{hypcap}    

\begin{document}

\cen{ {\Large {\bfseries Development of a defacing algorithm to protect the privacy of head and neck cancer patients in publicly-accessible radiotherapy datasets }} \\  
\vspace*{10mm}
{\large
 Kayla O’Sullivan-Steben$^1$, Luc Galarneau$^{1-2}$, John Kildea$^{1-3}$} \\
 {\small \textit{
$^1$ Medical Physics Unit, McGill University,  Montreal, QC, Canada\\	
$^2$ Research Institute of the McGill University Health Centre, Montreal, QC, Canada\\
$^3$ Gerald Bronfman Department of Oncology, McGill University, Montreal, QC, Canada}
}
\vspace{5mm} 
\\\today
}

\pagenumbering{roman}
\setcounter{page}{1}
\pagestyle{plain}
\vspace{-7mm}

\cen {Corresponding author email: kayla.osullivan-steben@mail.mcgill.ca}


\begin{abstract}
\noindent {\bf Background:} The increase in public medical imaging datasets has raised concerns about potential patient reidentification from head CT scans. However, existing defacing algorithms, which help protect patient confidentiality, fail  to preserve critical radiotherapy structures, including Organs at Risk (OARs) and Planning Target Volumes (PTVs) in head and neck cancer (HNC) patients. Furthermore, current algorithms do not address the defacing of DICOM-RT Structure Set and Dose data, which also contain information for facial surface rendering.\\ 
{\bf Purpose:} To develop and validate a novel automated defacing algorithm that preserves OARs and PTVs while removing identifiable features from HNC CTs and DICOM-RT data.\\
{\bf Methods:} Eye contours were used as landmarks to automate the removal of CT pixels above the inferior-most slice of the eye and anterior to the midpoint of the eye. Pixels within PTVs were retained if they intersected with the removed region. The body contour and dose map were then reshaped to reflect the defaced image. We validated our approach on 829 HNC CT-simulation scans from 622 patients. To evaluate privacy protection, we applied the FaceNet512 facial recognition algorithm before and after defacing on 3D-rendered CT scan pairs from 70 patients at two time points. To assess research utility, we examined the impact of defacing on auto-contouring performance using LimbusAI and analyzed the locations of PTVs relative to the defaced regions.\\
{\bf Results:} Before defacing, the facial recognition algorithm matched 97\% of patients' CT scans. After defacing, this rate dropped to just 4\%. LimbusAI effectively auto-contoured organs in the defaced CTs, with perfect Dice scores of 1 for OARs below the defaced region, and excellent scores exceeding 0.95 for OARs on the same slices as the defaced region. PTV analysis revealed that 86\% of PTVs were entirely below the cropped region, 9.1\% were on the same slice as the crop without overlap, and only 4.9\% extended into the cropped area. All overlapping PTVs were preserved through our algorithm's design.
 \\
{\bf Conclusions:} We developed a novel defacing algorithm that anonymizes HNC CT scans and related DICOM-RT data. Our algorithm balances patient privacy while preserving essential structures for radiotherapy research, facilitating the sharing of HNC imaging datasets for Big Data and AI.
 \\

\end{abstract}

\newpage     


\newpage

\setlength{\baselineskip}{0.7cm}      

\pagenumbering{arabic}
\setcounter{page}{1}
\pagestyle{fancy}
\section{Introduction}
The proliferation of medical imaging datasets in the public domain has raised concerns about potential patient reidentification from CT scans of the head\cite{parker_canadian_2021,parks_automated_2017,mazura_facial_2012}. Various defacing algorithms have been produced to mitigate such concerns. However, these tools overlook the need to preserve critical structures that are important for radiotherapy research\cite{wahid_artificial_2022, sahlsten_segmentation_2023}. Therefore, we developed a novel automated defacing algorithm that preserves Organs at Risk (OARs) and Planning Target Volumes (PTVs) for radiotherapy planning while removing identifiable features from head and neck cancer (HNC) CT scans and associated DICOM-RT data.

\subsection{Background}
In the present era of Big Data and Artificial Intelligence, there is an increasing demand for publicly accessible imaging datasets for radiation oncology research. When publishing datasets, it is crucial to remove any identifying information to protect patient privacy\cite{parker_canadian_2021}. However, this task becomes particularly complex in the case of HNC patients, as the body surface renderings of their 3D image scans could potentially be used for facial recognition and re-identification\cite{parker_canadian_2021,parks_automated_2017,mazura_facial_2012}. For example, Schwarz and colleagues (2022)\cite{schwarz_face_2022} found that surface renderings of MRI scans could be matched to a photo of the same patient with 97-98\% accuracy, while CTs were matched at 78\% accuracy using an automated facial recognition tool. Therefore, to enable more researchers to contribute without ethical concerns to public datasets for AI and Big Data applications in HNC, it is essential that the community has access to robust de-identification techniques \textit{Defacing}, in particular, contributes to data anonymization by obscuring identifiable facial features in imaging datasets.

\subsection{Existing defacing algorithms}
While defacing is commonly used to address privacy concerns, existing defacing algorithms were developed primarily for neuroimaging research and do not necessarily suit the needs of radiotherapy-related studies\cite{sahlsten_segmentation_2023}. More specifically, these tools typically focus on preserving brain structures but fail to consider other critical structures, such as the many OARs and PTVs in the CT-simulation (CT-sim) and/or cone-beam CT scans of the head and neck region used for radiotherapy delivery. As a result, critical structures can become distorted or removed entirely when conventional defacing techniques are used. This problem was illustrated by Wahid et al. (2022)\cite{wahid_artificial_2022}, showing how four state-of-the-art defacing algorithms obscure or remove important HNC OARs like the lymph node levels and salivary glands. 

Maintaining the integrity of such structures is crucial for radiotherapy research, which can involve tasks such as auto-segmentation, radiomics, and tracking of tumour volumes and anatomical changes\cite{volpe_machine_2021}. For instance, Sahlsten et al (2023)\cite{sahlsten_segmentation_2023} demonstrated how current defacing algorithms impede HNC auto-segmentation research. In their study, they examined eight publicly available defacing algorithms and found that five were incapable of defacing CT images, as they were specifically designed for MRI use. The remaining three tools did deface the images, but caused a significant decline in performance of their auto-segmentation algorithm when it was trained and tested on defaced CTs compared to the original CTs. 

As an additional but important consideration in the radiotherapy domain, conventional defacing algorithms do not consider the anonymization of DICOM-RT Structure Set and Dose data that radiotherapy treatment plans are stored in. These data also contain 3D anatomical information that can be used for the surface rendering of a patient’s face and so must also be defaced.

\subsection{Our approach}
In this work, we aimed to develop an automated defacing algorithm for HNC CT scans that preserves OARs and PTVs while removing identifiable features like the eyes, eyebrows, and forehead. It was also important that our technique extends to defacing DICOM-RT Structure Set and Dose data to ensure an added layer of privacy protection for radiotherapy patients in addition to de-identification. We validated our defacing algorithm by comparing the performance of facial recognition and auto-segmentation algorithms before and after defacing, and by examining the location of HNC tumours relative to the defaced area. We believe that this work can facilitate the safe sharing of HNC imaging datasets by providing a method to anonymize CT images while maintaining their utility for radiotherapy research. To the best of our knowledge, this is the first defacing algorithm designed specifically for HNC radiotherapy data.


\section{Methods}
This work was carried out on a retrospective single-centre patient dataset of 622 HNC patients who underwent radiotherapy treatment between January 1, 2017 and March 31st, 2024. In total, the dataset comprised 829 CT-sim scans along with their associated Structure Sets and Dose maps. The protocol for this retrospective research study was approved by the Research Ethics Board (REB) of the McGill University Health Centre [project number 2025-11285]. All work of the study was conducted in accordance with the Canada Tri-Council Policy Statement: Ethical Conduct for Research Involving Humans (TCPS 2). Additionally, all potentially-identifying fields of the CT and DICOM-RT data were anonymized by the Eclipse Treatment Planning System (Varian Medical Systems, Inc. Palo Alto, CA, USA) on export, thus stripping them of any identifiers such as names, dates, etc. 

\subsection{Selecting the region to deface}

In our study, we aimed to strike a balance between ensuring the anonymity of the images and maintaining their utility for radiotherapy research. To achieve this, we conducted a needs assessment, summarized in Table \ref{tab:Needs_Assessment}, which outlines the key considerations and proposed solutions that guided the algorithm’s development.  The resulting approach removes the region anterior to the centre of the eye and superior to the bottom of the eyes, while preserving anatomical information related to the PTV, even when located within the defaced area.

\begin{table}[ht!]
\begin{center}
\captionv{10}{}{Summary of the needs assessment for our defacing algorithm.
\label{tab:Needs_Assessment}
\vspace*{1ex}
}

\begin{tabular}{|p{4cm}|p{5.6cm}|p{5.6cm}|}
\hline
\textbf{Need}                                           & \textbf{Consideration}                         & \textbf{Specific Solution}                                                                                                                 \\ \hline
1) Protect patient privacy                                 & Facial recognition algorithms typically rely on facial landmarks, such as the eyes, eyebrows, forehead, nose, and mouth\cite{adjabi_past_2020}.                                                                               & Remove facial landmarks used by facial recognition algorithms.                                                                                                                            
\\
\cline{1-3}
2) Preserve tumour volumes for HNC research.               & HNC predominantly occurs in the oral cavity, larynx, and pharynx\cite{barsouk_epidemiology_2023}.                                                                                 & Retain the inferior portion of the head starting at least at the oral cavity; preserve all PTV pixels, even if within the proposed cropped region.                                                    \\ \cline{1-3}
3) Preserve OARs for radiotherapy research.                & The OARs near the surface of the face–which are most susceptible to being removed during defacing–include: eyes, oral cavity, lips, mandible, lymph nodes, submandibular and parotid glands. & Retain the inferior portion of the head, ensuring removal of only those structures that compromise privacy.                                                           \\ \cline{1-3}
4) Ensure utility for neuroimaging studies.                & The brain structure should remain intact in the image.                                                                     & Remove only pixels anterior to the centre of the eye.                       \\ \cline{1-3}
5) Ensure utility for HNC radiotherapy replanning studies. & Replanning often hinges on subtle external changes such as weight loss and local anatomical changes.                                                                                                      & Avoid deformation or removal of the skin surface around the tumour, chin, and neck.                                                                               \\ \hline
\end{tabular}
\end{center}
\end{table}

\subsection{Automated defacing workflow}
\begin{figure}[ht!]
   \begin{center}
   \includegraphics[width=8cm]{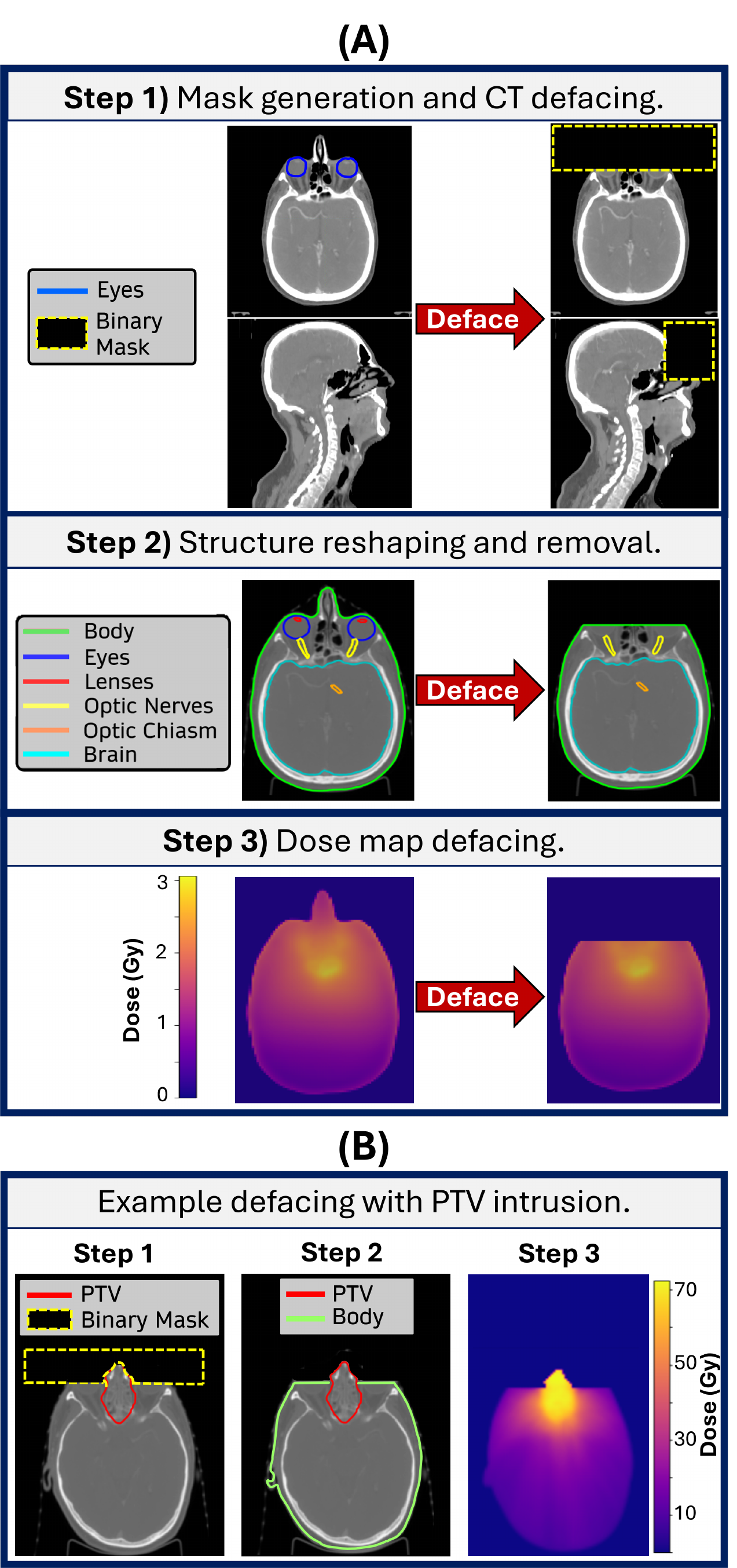}
   \captionv{12}{}
   {(A) Overview of the automated defacing algorithm’s workflow. (B) Example of the workflow applied to a patient whose PTV intrudes into the defaced region.
   \label{fig:Defacing_Workflow} 
    }  
    \end{center}
\end{figure}

An overview of our automated defacing workflow is presented in Figure \ref{fig:Defacing_Workflow}. In the first step, the eye contours–as delineated by the dosimetrists/radiation oncologists–are extracted from the DICOM-RT Structure Set data of the CT-sim image. Starting from the inferior-most slice of the eye structure and moving toward the top of the head, the algorithm generates a binary mask (values of 0 and 1) that removes all pixels anterior to the centre point of the eye contour from each image slice of the CT-sim. The algorithm then checks for any PTV or brain structures that overlap with the initially cropped area, and, if applicable, modifies the mask to retain the image pixels corresponding to the PTV and brain structures to preserve the target and organ anatomy. The mask is then applied to the image to remove the defaced pixels, and can likewise be applied to any other images (e.g. daily cone-beam CTs) that have been registered to the simulation-CT.

In the second step, the array of x,y,z-coordinates of the body contour are updated to reflect the modified cropped images. Additionally, the eye, lens, and cornea contours are removed from the Structure Set. The algorithm then checks whether any other contours (aside from the PTV and brain) protrude into the cropped region and, if so, reshapes them accordingly. The PTV contours are kept in the Structure Set, even when they overlap with the cropped region. Finally, in the third step, the mask is resized and resampled to the size and spacing of the Dose map contained in the RT Dose file. The new mask is then applied to the Dose map to deface it. Once again, in cases where the PTV overlaps with the cropped region, the mask also retains the Dose map voxels corresponding to the overlapping PTV image pixels.

To test our workflow, we ran the automated algorithm over a HNC dataset of 829 unique HNC CT-sim scans corresponding to 622 patients.  We visually inspected each defaced CT, Structure Set and Dose map to ensure proper defacing. Additionally, we calculated the average running time required to deface one CT and associated DICOM-RT data to assess computational feasibility. The defacing code was implemented in Python (version 3.8) and is available in the following GitHub repository: https://github.com/kildealab/defaceRT.

\subsection{Validation - patient privacy}
To evaluate privacy protection, we conducted facial recognition tests on 2D images (screenshots) of 3D renders of the CT scans before and after defacing. This methodology is consistent with other studies in the literature investigating facial recognition using CTs\cite{parks_automated_2017,mazura_facial_2012,uchida_-identification_2023}. We performed these tests on a randomly selected subset of 70 patients from our dataset, each of whom had a second independent CT-sim scan available for matching (taken for replanning purposes, typically a week or more after the initial CT). Using two CT scans taken at different time points is essential for replicating a real-life facial recognition scenario. It is akin to comparing two different photos of the same person, whereas comparing two photos or CTs taken in close temporal proximity could yield better (and potentially misleading) matching results. 

\subsubsection{Image preparation}
To obtain images of the 3D rendered faces for facial recognition, the body contours of the CTs were rendered in 3D in the Eclipse Treatment Planning System (Varian Medical Systems, Inc. Palo Alto, CA, USA). All body contours were set to the same white colour and a 2D screenshot was taken of the render facing forward. This process was repeated to provide three CT scans for each patient: one at the first time point ($CT_{t1}$), one independent scan at the second time point ($CT_{t2}$), and one defaced scan corresponding to the first time point ($dCT_{t1}$).

\subsubsection{Face detection and recognition}
Facial recognition algorithms typically comprise three main steps, which we implemented using the open-source DeepFace library\cite{serengil_lightface_2020} (github.com/serengil/deepface). First, a detector model locates the face within an image so that it can be isolated for recognition. In our case, we tested all facial detection methods available in DeepFace and opted to use RetinaFace\cite{deng_retinaface_2019}, as it successfully detected all 140 faces (70 pairs) in our non-defaced CT scans. Next, a recognition model transforms these detected faces into vector embeddings. For this step, we employed the FaceNet512 algorithm\cite{schroff_facenet_2015}, as it has been shown to outperform other existing publicly available models in facial recognition tasks\cite{firmansyah_comparison_2023,serengil_benchmark_2024} and has been used in other CT imaging defacing studies\cite{selfridge_facial_2023,mahmutoglu_deep_2024}. In the final step, a pair of facial embeddings are compared using a distance metric, with closer embeddings (smaller distances) indicating higher similarity and thus a greater likelihood that they represent the same person. For this comparison, we used DeepFace’s default cosine distance metric.

\begin{figure}[ht!]

   \begin{center}
   \vspace{4ex}
   \includegraphics[width=8cm]{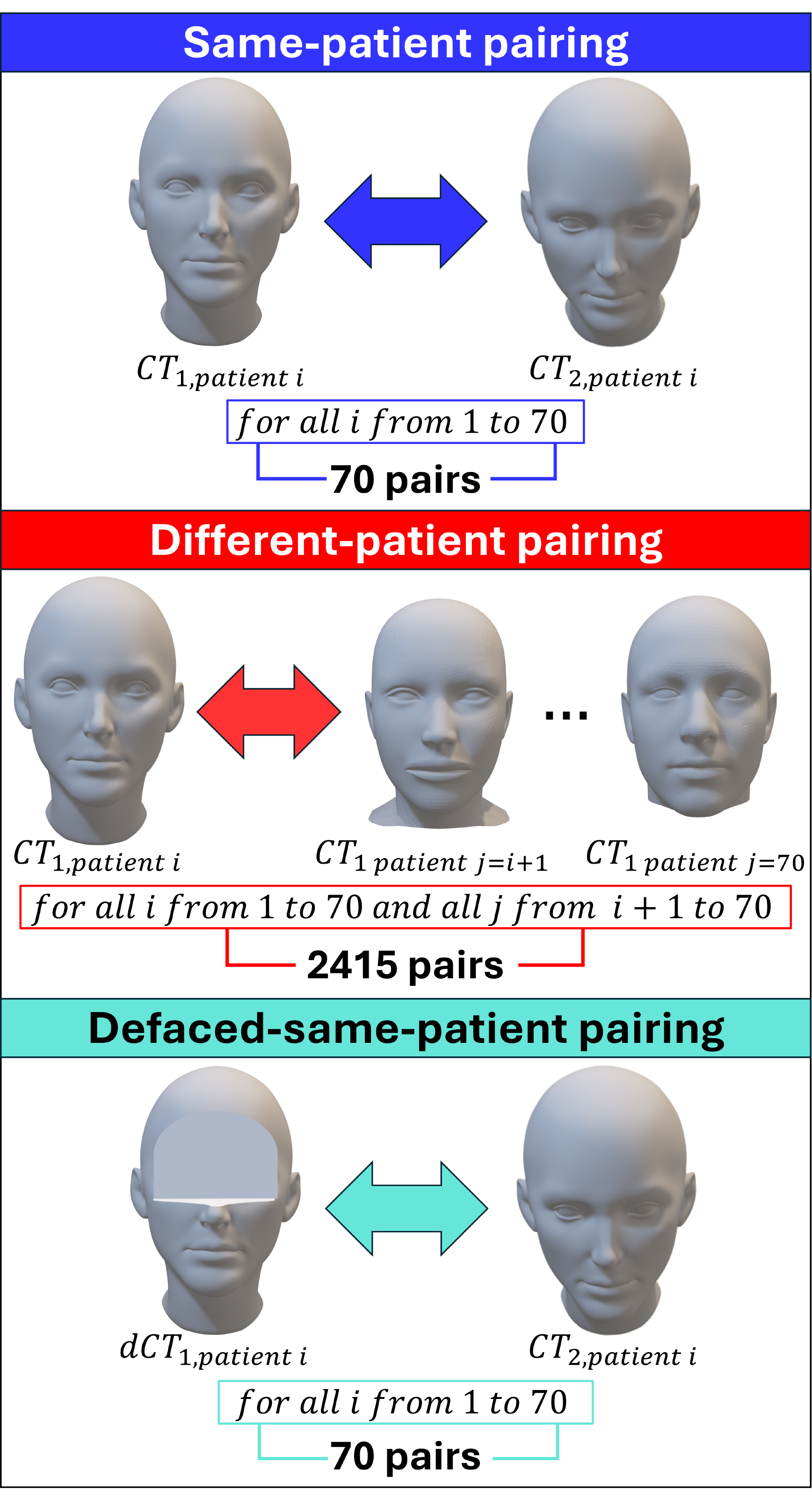}
   \captionv{12}{}
   {Overview of the three groups of facial recognition tests performed. Each pair of image comparisons yields one cosine distance. Note that these images are artist-rendered 3D faces for visualization purposes. They do not represent the CTs of real patients.
   \label{fig:Comparison_Groups} 
    } 
    \end{center}
\end{figure}

We performed facial recognition tests and obtained cosine distances for the following three comparison pairing groups, each of which is visualized in Figure \ref{fig:Comparison_Groups}: 
\begin{enumerate}
    \item \textbf{Same-patient pairing}: comparison between CT scans of the same patient at two different time points ($CT_{t1, patient\ i}$ vs $CT_{t2, patient \ i}$ for all $i$).
    \item \textbf{Different-patient pairing}: comparison between CTs of different patients (($CT_{t1,patient\ i}$ vs $CT_{t1, patient\ j}$ for all $j = i+1$ to $70$) for all $i$).
    \item \textbf{Defaced-same-patient pairing}: comparison between a defaced CT from the first time point and the original CT from the second time point of the same patient ($dCT_{t1, patient\ i}$ vs $CT_{t2, patient\ i}$ for all $i$).
\end{enumerate}

\subsubsection{Baselining exercise}
Given that FaceNet512 was originally developed for 2D photographic images and was not explicitly trained on 2D renderings of 3D images, it was important to establish a baseline to ensure that the model can adequately distinguish between two CT scans of the same patient (expected low cosine distances) and two CT scans of different patients (expected high cosine distances). Although the aforementioned studies in the literature\cite{selfridge_facial_2023,mahmutoglu_deep_2024} did not undertake this additional step, we did so to satisfy ourselves that FaceNet512 is an appropriate algorithm to use for our use case involving radiotherapy CT-sim scans. 

We performed the Mann-Whitney U test to confirm that the set of cosine distances obtained for the same-patient pairing group and the different-patient pairing group were statistically different. Good differentiation in cosine distances between these two pairing groups would indicate that the FaceNet512 algorithm is capable of distinguishing same patients and different patients.  

The optimal cosine distance threshold for separating the two pairing groups was determined by maximizing the Youden Index\cite{youden_index_1950} using scikit-learn's ROC curve implementation\cite{pedregosa_scikit-learn_2011}, which measures the tradeoff between the true positive rate and false positive rate. Using this threshold, we determined the number of correctly matched patients before defacing, as well as the number of false positive matches when comparing CTs of different patients.

\subsubsection{Defacing evaluation}
With the baselining completed, we compared the cosine distances obtained for the defaced-same-patient pairing group with the baseline groups of the same-patient and different-patient pairing groups. We hypothesized that the defaced images would produce cosine distances that are consistent with those of the different-patient pairing group, indicating successful anonymization. Using the previously defined threshold, we determined the number of correctly matched patients after defacing.

We conducted a Wilcoxon signed-rank test to determine if there was a significant difference between the cosine distances for the same patient before and after defacing. Finally, we performed the Mann-Whitney U test to compare the cosine distances of the defaced-same-patient pairing and different-patient pairing groups to determine if these two groups are distinguishable. 

A summary of the statistical comparisons and expected outcomes for both the baselining exercise and the defacing evaluation are presented in Table \ref{tab:Stats_Tests}. For all statistical tests, a p-value $<$ 0.005 was considered significant.

\vspace{1ex}
\begin{table}[ht]
\begin{center}
\captionv{10}{}{Statistical comparison tests and expected outcomes of cosine distances for the different pairing groups.}
\label{tab:Stats_Tests}
\begin{tabular}{|c|c|c|c|}
\hline
\textbf{Test Purpose}                                                           & \textbf{\begin{tabular}[c]{@{}c@{}}Cosine Distance \\ Comparison Group\end{tabular}}                                            & \textbf{Test Name}                                              & \textbf{\begin{tabular}[c]{@{}c@{}}Expected \\ Outcome\end{tabular}} \\ \hline
Baselining                                                                      & \begin{tabular}[c]{@{}c@{}}Same-patient pairing \\ (70 pairs) vs Different-patient \\ pairing (2415 pairs)\end{tabular}         & Mann-Whittney U                                                 & \begin{tabular}[c]{@{}c@{}}Significantly \\ different\end{tabular}      \\ \hline
\multirow{2}{*}{\begin{tabular}[c]{@{}c@{}}Defacing \\ Evaluation\end{tabular}} & \begin{tabular}[c]{@{}c@{}}Same-patient pairing \\ (70 pairs) vs Defaced-same-\\ patient pairing (70 pairs)\end{tabular}        & \begin{tabular}[c]{@{}c@{}}Wilcoxon \\ Signed-Rank\end{tabular} & \begin{tabular}[c]{@{}c@{}}Significantly \\ different\end{tabular}      \\ \cline{2-4} 
& \begin{tabular}[c]{@{}c@{}}Different-patient pairing \\ (2415 pairs) vs Defaced-same-\\ patient pairing (70 pairs)\end{tabular} & Mann-Whittney U                                                 & Indistinguishable                                                       \\ \hline
\end{tabular}
\end{center}
\end{table}

\subsection{Utility for radiotherapy research}
To assess the utility of our defaced images for radiotherapy research, we investigated the effects the defacing had on the two main structure types of interest for radiotherapy research: OARs and PTVs. 

\subsubsection{OARs}
We used LimbusAI (Limbus AI Inc, Regina, SK, Canada), the auto-segmentation software used in our clinic, to automatically generate contours for head and neck OARs on both the original and defaced CTs of the same subset of 70 patients. Contouring was performed for all 46 OARs defined in our clinic’s ‘Head and Neck’ LimbusAI template.

For each patient, we then calculated the Dice coefficient to measure the degree of overlap between the contours on the original and defaced CTs for each OAR contoured. For a 3D segmentation of a given OAR on the original CT ($X$) and defaced CT ($Y$), the Dice coefficient ($DSC$) is defined as:
\[DSC = \frac{2|X\cap Y|}{|X|+|Y|}\]
A result of 1 indicates that the two volumes overlap completely, whereas a result of 0 indicates that they do not overlap at all. The intent of calculating these Dice scores was to allow us to examine to what extent each OAR and its surrounding tissues can still be segmented and analysed in radiotherapy research using the defaced CTs.

\subsubsection{PTVs}
We analyzed the locations of the PTVs relative to the removed facial regions in our complete dataset of 622 patients. Specifically, we quantified the fractions of PTVs that fell into one of three categories: (1) completely below the cropped region, (2) on the same slices as the cropped region, but not overlapping it, and (3) overlapping with the cropped region. This assessment provided insight into the potential impact of the defacing process on tumour regions, as PTVs located in or on the same slice as the removed region may encounter additional research limitations compared to those completely below it.

\section{Results}
\subsection{Real-world defacing}

\begin{figure}[ht]
   \begin{center}
   \includegraphics[width=8cm]{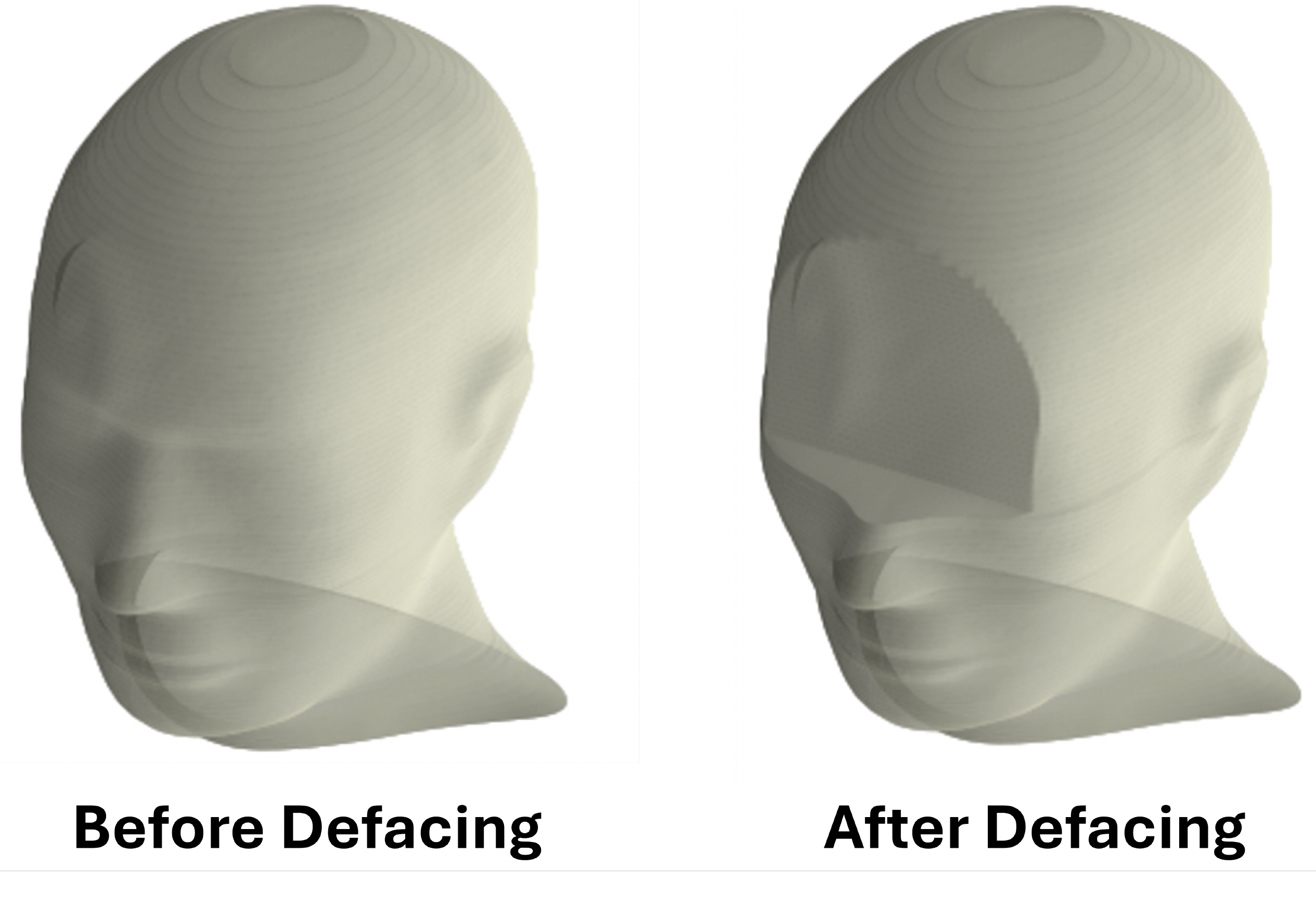}
   \captionv{12}{}
   {3D surface rendering of the reconstructed face of a head phantom before and after defacing. The head phantom data were retrieved from the SlicerRtData GitHub repository\cite{noauthor_slicerrtdataeclipse-8120-phantom-ent_nodate}
   \label{fig:Before_After_Defacing} 
    } 
    \end{center}
\end{figure}

Our algorithm was able to automatically deface 793 (96\%) of the 829 CTs and associated Structure Sets and Dose maps in our dataset. The 36 (4\%) scans that were not successfully automatically defaced did not have eyes contoured, and thus would require an additional step of either manual or automated contouring before defacing. To best represent real-world conditions, we did not undertake the additional contouring step. All subsequent analyses were thus performed on the 793 automatically defaced CT scans. The code executed with a mean runtime of 13 ± 6 seconds per CT scan (including defacing of the Structure Set and Dose map) on a machine equipped with a virtual Intel Core Processor (Skylake, IBRS) and 8 GB of RAM. For illustration purposes, Figure \ref{fig:Before_After_Defacing} shows an example of a public-domain\cite{noauthor_slicerrtdataeclipse-8120-phantom-ent_nodate} 3D rendered face before and after defacing.

\subsection{Privacy evaluation}
\begin{figure}[ht]
   \begin{center}
   \includegraphics[width=16cm]{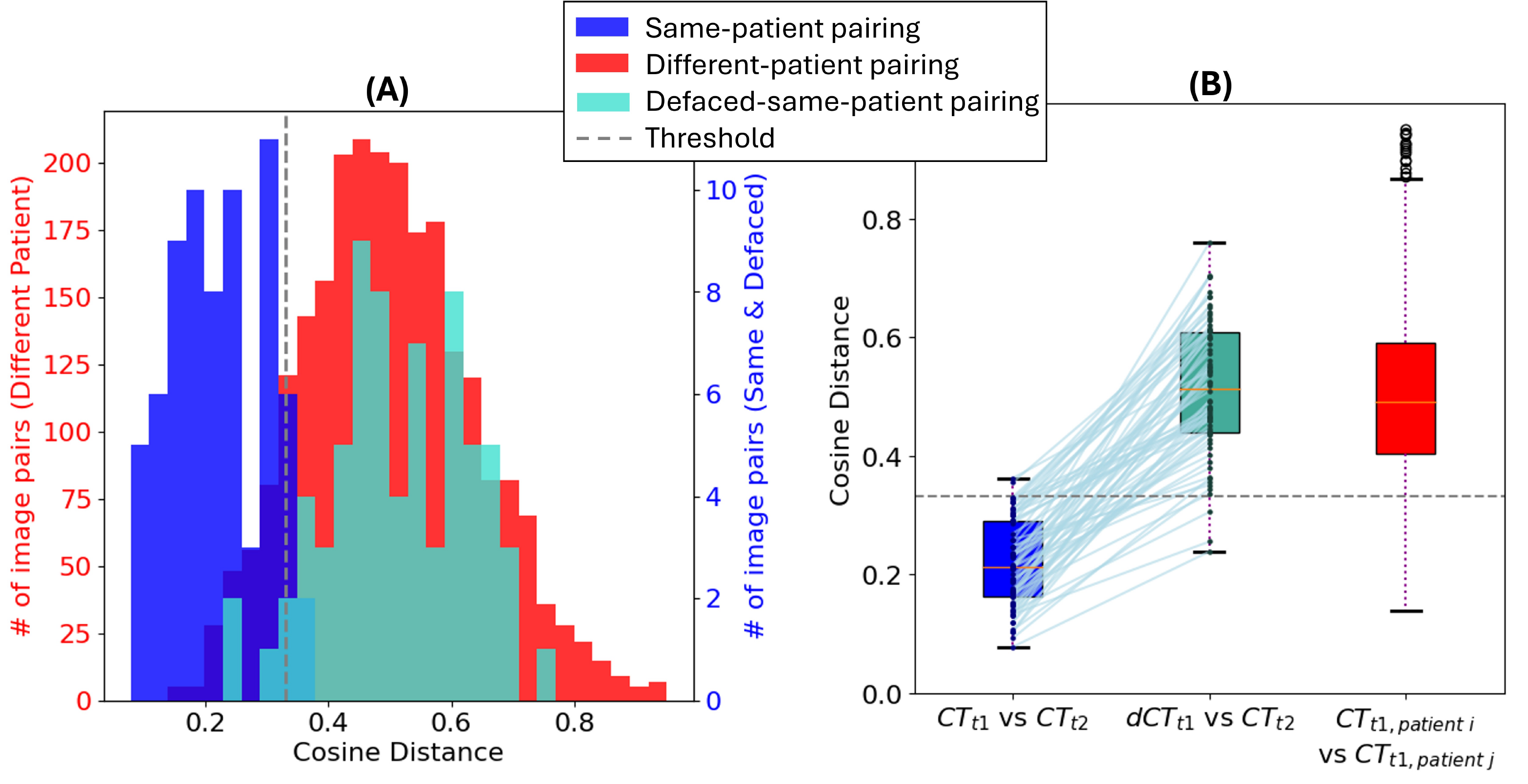}
   \captionv{12}{}
   {Results of the FaceNet512 facial recognition algorithm on 70 patients for our three pairing groups. Lower cosine distances indicate a higher likelihood that two scans are from the same patient. (A) presents histograms of the cosine distances of the three pairing groups tested. (B) shows the same data presented in whisker plots, with blue lines connecting data for the same patient. 

   \label{fig:Facial_Recognition_Results} 
    } 
    \end{center}
\end{figure}

Figure \ref{fig:Facial_Recognition_Results}(A) presents histograms of the cosine distances for the three pairing groups tested. The cosine distances for the same-patient pairing group ($CT_{t1, patient\ i}$ vs $CT_{t2, patient\ i}$) were significantly different from the cosine distances for the different-patient pairing group ($CT_{t1,patient\ i}$ vs $CT_{t1, patient\ j}$), with a p-value of $5.85^{-40}$. This significant difference indicates that FaceNet512 can reliably distinguish the renderings of CT scans of the same patient from CT scans of different patients. After defacing, we found that the cosine distances of the defaced-same-patient pairing group  ($dCT_{t1, patient\ i}$ vs $CT_{t2, patient\ i}$) were significantly different from the cosine distances of the same-patient pairing group ($CT_{t1, patient\ i}$ vs $CT_{t2, patient\ i}$), with a p-value of $1.60^{-22}$. Furthermore, the cosine distances for the defaced-same-patient pairing group were statistically indistinguishable from the cosine distances of different-patient pairing group (p-value of 0.40).

Using the maximized Youden Index, the threshold cosine distance was determined to be 0.331. Using this value, the baseline match rate for same-patient pairs (i.e. before defacing) was 97\% (68/70) with a false positive rate of 11\% (258/2415). After defacing, the match rate decreased to 4\% (3/70), which is notably lower than the false positive rate. Figure \ref{fig:Facial_Recognition_Results}(B) illustrates the increase in cosine distances (i.e. decrease in match likelihood) between each pair of CTs before and after defacing.

\subsection{Evaluation of utility for radiotherapy research}
\subsubsection{OARs}
\begin{figure}[ht]
   \begin{center}
   \includegraphics[width=10cm]{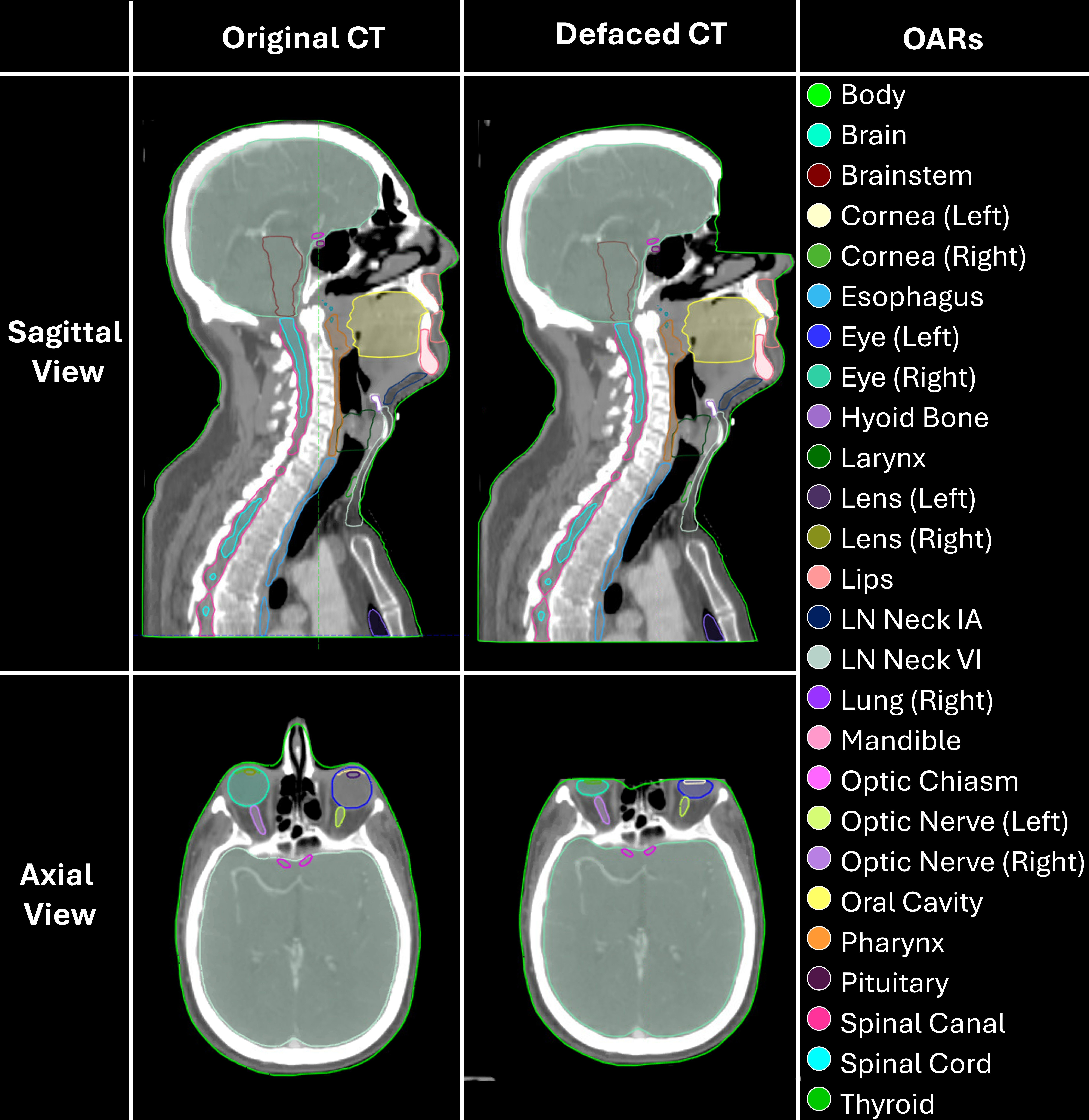}
   \captionv{12}{}
   {Visualization of LimbusAI’s auto-contoured OARs before and after defacing on a sample patient. Not pictured are the brachial plexuses, clavicles, cochleas, hippocampi, left lung, submandibular glands, parotid glands, and the following right and left Lymph Node (LN) levels: Neck, Neck 2347AB, Neck IB, and Neck V. 
   \label{fig:OARs_Before_After_Defacing} 
    } 
    \end{center}
\end{figure}

The LimbusAI software was able to generate contours on all of the 70 original and 70 defaced CT scans. A visualization of the contours before and after defacing on a sample patient are provided in Figure \ref{fig:OARs_Before_After_Defacing}.

\begin{figure}[ht]
   \begin{center}
   \includegraphics[width=16cm]{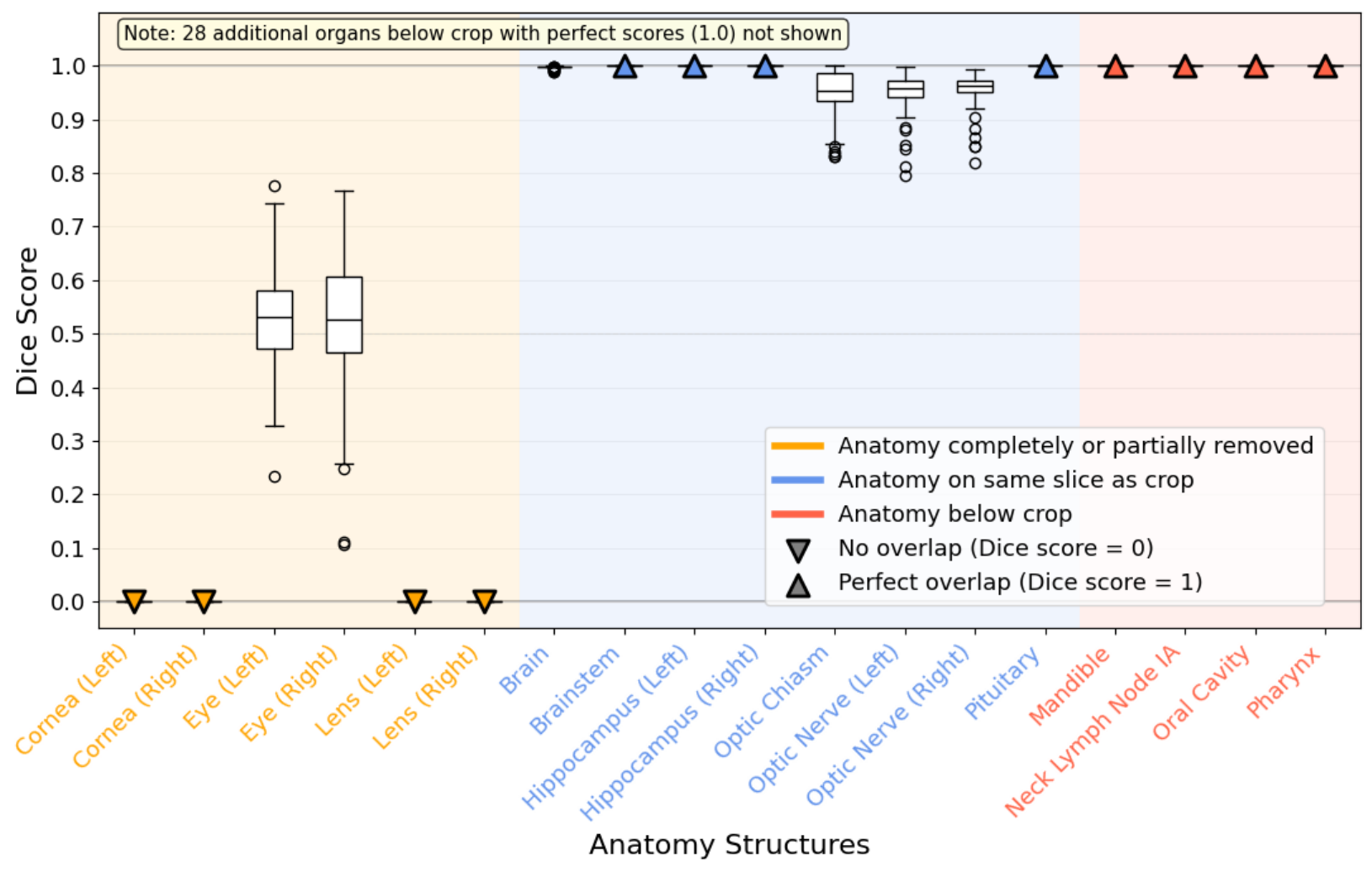 }
   \captionv{12}{}
   {Dice scores measuring overlap between auto-contoured OARs on the defaced CTs and the original CTs for 70 patients. A value of 1 indicates that the contours were identical on the original and defaced CTs. Additional OARs below the cropped region are not shown in this graph. They include brachial plexuses, clavicles, cochleas, esophagus, hyoid bone, larynx, lips,lungs, neck lymph node levels, parotid glands, spinal canal, spinal cord, submandibular glands, and thyroid, each of which had a perfect Dice score of 1 ($SD=0$). 

   \label{fig:Dice_Scores} 
    } 
    \end{center}
\end{figure}

The Dice scores comparing each of the 46 auto-contoured OARs on the original and defaced CT scans are presented in Figure \ref{fig:Dice_Scores}. As expected, the six anatomical structures that were partially or completely removed exhibited low Dice scores, indicating poor overlap. Specifically, the lenses and corneas had Dice scores of 0 ($SD=0$), while the left and right eyes had Dice scores of 0.5 ($SD=0.1$). These results are consistent with the fact that the corneas and lenses were entirely removed, and about half of the eyes were removed.

For the eight anatomical structures located on the same slice as the cropped region, but not overlapping it, Dice scores were all close to or equal to 1, indicating almost perfect overlap. Specifically, the brain, optic chiasm, left optic nerve, and right optic nerve had mean Dice scores of 0.999 ($SD=0.002$), 0.95 ($SD=0.05$), 0.95 ($SD=0.04$), and  0.96 ($SD=0.03$), respectively. The left and right hippocampi, brainstem, and pituitary all had perfect Dice scores of 1 ($SD = 0$).

Finally, all of the 32 OARs inferior to the cropped region were auto-contoured identically in the defaced and original CTs, yielding perfect Dice scores of 1 ($SD = 0$). Notably, this includes the 10 lymph node levels, which are routinely used by radiation oncologists to delineate Clinical Target Volumes (CTVs).

\subsubsection{PTVs}
We found that for 86.0\% (682/793) of the CT scans that were successfully defaced, the PTVs were located entirely inferior to the cropped region, indicating that the beam entry points were unaffected by the defacing process for the majority of patients. 9.1\% (72/793) of CTs had a PTV that was partially on the same slice as the defaced region, but not overlapping the cropped pixels. Only 4.9\% (39/793) of the CT scans had a PTV that overlapped with the initial cropped region. These cases corresponded to patients with diagnoses in the nasal cavity (22/39), the sinuses (11/39), the palate (3/39), the cheek (2/39), and the nasopharynx (1/39).

\section{Discussion}

In this study, we developed a novel defacing algorithm for HNC CT scans and associated DICOM-RT data. Our algorithm automatically removes identifiable facial features – namely the eyes, eyebrows and forehead – while preserving critical anatomical structures needed for radiotherapy research. Additionally, our algorithm is fast and requires minimal computational power. To our knowledge, this is the first implementation of a defacing algorithm specifically designed for HNC CT data that also includes the defacing of Structure Sets and Dose maps, addressing an important privacy vulnerability in radiotherapy data sharing.

Our defacing algorithm successfully addressed all identified privacy concerns, with facial recognition rates (between defaced and non-defaced images) decreasing from 97\% to 4\% following defacing. These results align with similar studies, such as Schwarz et al (2022)\cite{schwarz_face_2022} and Selfridge et al (2023)\cite{selfridge_facial_2023}, who reported decreases from 78\% to 5\% and 93\% to 7\% with their respective defacing algorithms on CT scans but without radiotherapy considerations. As such, our approach achieves comparable privacy to existing algorithms, while offering additional advantages for radiotherapy research.

A key advantage of our approach is the preservation of OARs essential for radiotherapy research. Previous studies have demonstrated that conventional defacing algorithms, which were primarily designed to preserve only the brain structure for neuroimaging studies, obscure or remove important HNC structures including lymph node levels and salivary glands\cite{wahid_artificial_2022,sahlsten_segmentation_2023}. This OAR degradation can have downstream impacts on research that uses these defaced images. For instance, Sahlsten et al. (2023)\cite{sahlsten_segmentation_2023} found that autosegmentation models trained on original CT scans performed poorly on defaced images, and models trained on defaced scans were less effective when tested on the original data.

In contrast, our defacing algorithm not only preserves these critical HNC structures by design, but also maintains sufficient surrounding anatomical context to enable accurate auto-segmentation. Our analysis revealed that auto-segmentation is virtually unaffected by defacing – aside from the intentionally removed structures – with perfect Dice scores of 1.0 for structures inferior to the cropped area and near-perfect scores ($>0.95$) for structures on the same slice as the crop.

Another advantage of our defacing algorithm is the systematic preservation of PTV structures. Not only are all PTV pixels retained in the image, but our evaluation confirmed that the majority of the PTVs (86.0\%) lie entirely below the cropped regions, with only 4.9\% extruding into the defaced area. To our knowledge, no other study has investigated the impact of defacing on PTVs, despite their critical value to radiotherapy research applications. However, given that other defacing algorithms remove or deform important OARs, it is reasonable to assume that PTVs would similarly be compromised by them.

While most existing defacing techniques were originally developed for MRI, a few more recent algorithms have been proposed specifically for CT scans. For example, Mahmutoglu et al (2024)\cite{mahmutoglu_deep_2024} and Lindholz et al (2025)\cite{lindholz_analyzing_2025} report deep-learning based defacing algorithms, but both remove a substantial portion of important HNC OARs around the mouth, resulting in the same limitation of many of the aforementioned MRI-focused defacing techniques.

Rather than removing pixels entirely, some researchers have explored deformation and blurring methods that attempt to preserve facial resemblance. Uchida et al (2023)\cite{uchida_-identification_2023} proposed a deformation-based de-identification method that manipulates head CT Images according to 400 control points set on the surface rendering of the patient’s face. This approach maintains facial resemblance, but the process is manual and thus not feasible for large-scale datasets. Selfridge et al (2023)\cite{selfridge_facial_2023} proposed a blurring method that, although automatic, severely deforms the facial structure. Importantly, both studies only investigated the effect of defacing on the brain structure, while appearing to deform many of the OARs near the anterior of the head. These anatomical distortions can interfere with radiotherapy applications requiring precise anatomical measurements, such as adaptive HNC treatment planning studies that rely on tracking body shrinkage and skin separation around the lower face and neck. By comparison, our pixel-removal approach maintains clean undeformed anatomical boundaries in the regions below the defaced region.

Overall, in addressing the limitations of current defacing techniques, our algorithm supports a broad spectrum of research applications in radiotherapy that rely on anatomical structural information, such as autosegmentation model development, radiomics analyses, dosimetric studies, anatomical change modeling, tumour volume tracking, and treatment planning.

\noindent{\textbf{Limitations}}\\
Our study has several limitations. First, the algorithm relies on pre-contoured eye structures for radiotherapy treatment planning, which were absent in about 4\% of our dataset. However, given the widespread availability of auto-contouring tools and the ease of contouring the eyes, this limitation can be relatively easily overcome. 

Second, for the small subset of patients (4.9\% in our dataset) with PTVs extending into the defaced region (primarily those with nasal cavity and sinus tumours), comprehensive dosimetric studies may be limited since nearby OARs like the eyes and lenses are of higher dosimetric importance in these cases. For these cases, further investigations are required to properly quantify the dosimetric impact of defacing.

Third, although we demonstrated that FaceNet512 provided a reasonable baseline for facial recognition using 2D renderings of 3D CT scans, it was not specifically trained to do so. However, to our knowledge, no other algorithms have been trained in such a context. As such, we believe that any publicly-available algorithm would suffer from the same limitation. Moreover, while our study focused on facial recognition between two CT renderings, real-world scenarios may involve comparisons between a CT rendering and publicly available photographs, which can be explored in future work.

Lastly, while our results indicate that our defacing algorithm substantially minimizes reidentification risks, there is always an inherent risk in sharing any healthcare data publicly. We also acknowledge the possibility that new facial recognition algorithms may be created in the future that are better at recognizing defaced patients. We therefore advise that all publicly shared head CT images – whether defaced or not – be distributed only via secure imaging archives with signed user agreements prohibiting the use of these images for non-research purposes, including 3D rendering for facial recognition.

\section{Conclusion}
In conclusion, we developed a defacing algorithm specifically for the defacing of HNC CT scans and their related DICOM-RT data. Our algorithm balances the need for patient privacy with the preservation of critical OARs and target structures that are crucial for radiotherapy research. By enabling the secure de-identification of imaging data while maintaining their research utility, this work addresses an important need in the era of Big Data and AI. Overall, this work can facilitate the sharing of HNC imaging datasets, which in turn can enable broader collaboration and accelerate advancements in radiotherapy research.

\section{Acknowledgments}
We gratefully acknowledge Victor Matassa for extracting the HNC patient IDs that met our inclusion criteria from the clinical database. We also thank Odette Rios-Ibacache, who, along with K.O., exported the CT-sim images and DICOM-RT data used in this project. K.O. acknowledges financial support from the Natural Sciences and Engineering Research Council (NSERC) of Canada, the Québec Ministère de la Santé et des Services Sociaux, and the CREATE Responsible Health and Healthcare Data Science (SDRDS) grant of NSERC. This work was also supported by the Fonds de rechereche du Québec–Santé dual-chair in AI and digital health held by J.K. and a research grant from the Rossy Cancer Network.

\section{Conflict of Interest Statement}
The authors have no relevant conflicts of interest to disclose.

\newpage
\section*{References}
\addcontentsline{toc}{section}{\numberline{}References}
\vspace*{-20mm}






\bibliographystyle{medphy.bst}    


\end{document}